\documentclass[%
 reprint,
 amsmath,amssymb,
 aps, prl,
 longbibliography
]{revtex4-1}

\usepackage{graphicx}
\usepackage{dcolumn}
\usepackage{bm}

\usepackage[plainpages=false,pdfpagelabels,colorlinks=true,linkcolor=black,urlcolor=black,citecolor=black]{hyperref}
\usepackage{braket}

\usepackage{color, soul}

\begin{document}

\newcommand{\Erec}{E_{\text{rec}}}
\newcommand{\trec}{t_{\text{rec}}}
\newcommand{\klat}{k_{\text{lat}}}
\newcommand{\llat}{\lambda_{\text{lat}}}
\newcommand{\Rb}{$^{87}$Rb}
\newcommand{\two}[2]{\begin{pmatrix}#1\\#2\end{pmatrix}}
\newcommand{\four}[4]{\begin{pmatrix}#1\\#2\\#3\\#4\end{pmatrix}}
\newcommand{\sm}{1 + \sqrt{2}}
\newcommand{\sqrttwo}{\frac{1}{\sqrt{2}}}
\newcommand{\Vd}{V^D(\vec{r})}
\newcommand{\ki}{\vec{\hat{G}}_i}
\newcommand{\kon}{\vec{\hat{G}}_1}
\newcommand{\ktw}{\vec{\hat{G}}_2}
\newcommand{\kth}{\vec{\hat{G}}_3}
\newcommand{\kfo}{\vec{\hat{G}}_4}
\newcommand{\bj}{\vec{b}_j}
\newcommand{\bi}{\vec{b}_i}
\newcommand{\lon}{\vec{\hat{l}}_1}
\newcommand{\ltw}{\vec{\hat{l}}_2}
\newcommand{\lth}{\vec{\hat{l}}_3}
\newcommand{\lfo}{\vec{\hat{l}}_4}
\renewcommand{\vec}[1]{\mathbf{#1}}


\title{Matter-wave diffraction from a quasicrystalline optical lattice}

\author{Konrad Viebahn}
\author{Matteo Sbroscia}
\author{Edward Carter}
\author{Jr-Chiun Yu}
\author{Ulrich Schneider}
\email{uws20@cam.ac.uk}
\affiliation{Cavendish Laboratory, University of Cambridge, J.~J.~Thomson Avenue, Cambridge CB3~0HE, United Kingdom}

\date{\today}

\begin{abstract}

  Quasicrystals are long-range ordered and yet non-periodic.
  This interplay results in a wealth of intriguing physical phenomena, such as the inheritance of topological properties from higher dimensions, and the presence of {non-trivial structure on all scales}.
  Here we report on the first experimental demonstration of an eightfold rotationally symmetric optical lattice, realising a two-dimensional quasicrystalline potential for ultracold atoms.
  Using matter-wave diffraction we observe the self-similarity of this quasicrystalline structure, in close analogy to the very first discovery of quasicrystals using electron diffraction.
  The diffraction dynamics on short timescales constitutes a continuous-time quantum walk on a homogeneous four-dimensional tight-binding lattice.
  These measurements pave the way for quantum simulations in fractal structures and higher dimensions.
  
\end{abstract}


\maketitle

Quasicrystals exhibit long-range order without being periodic~\cite{shechtman_metallic_1984, senechal_quasicrystals_1995, steurer_twenty_2004,
barber_aperiodic_2009, baake_aperiodic_2013, steurer_quasicrystals:_2018}.
Their long-range order manifests itself in sharp diffraction peaks, exactly as in their periodic counterparts.
However, diffraction patterns from quasicrystals often reveal rotational symmetries, most notably fivefold, eightfold, and tenfold, that are incompatible with translational symmetry. 
Therefore it immediately follows that long-range order in quasicrystals cannot originate from a periodic arrangement of unit cells but requires a different paradigm.
Quasicrystalline order naturally arises from an incommensurate projection of a higher-dimensional periodic lattice and thereby enables investigation of physics of higher dimensions, in particular in the context of topology~\cite{lang_edge_2012, kraus_topological_2012,kraus_topological_2012-1,kraus_four-dimensional_2013, matsuda_topological_2014}.
For instance, one-dimensional (1D) quasiperiodic models, such as the Fibonacci chain and the Aubry-Andre model, are closely connected to the celebrated two-dimensional (2D) Harper-Hofstadter model, and inherit their topologically protected edge states~\cite{kraus_topological_2012-1, matsuda_topological_2014}.
An alternative approach to constructing quasicrystals was described by Penrose~\cite{penrose_roaesthetics_1974} who discovered a set of tiles and associated matching rules that ensure aperiodic long-range order when tiling a plane~\cite{baake_aperiodic_2013}.
The resulting fivefold symmetric Penrose tiling and the closely related eightfold symmetric octagonal tiling~\cite{steurer_twenty_2004, baake_aperiodic_2013, chan_photonic_1998, jagannathan_eightfold_2013} (also known as Ammann-Beenker tiling) have become paradigms of 2D quasicrystals.
In addition to their disallowed rotational symmetries, these tilings have the remarkable feature of being self-similar in both real and reciprocal space~\cite{senechal_quasicrystals_1995, baake_aperiodic_2013}.
Self-similarity upon scaling in length by a certain factor (the silver mean $\sm$ in case of the octagonal tiling) implies that non-trivial structure is present on arbitrarily large scales.
Correspondingly, diffraction patterns from quasicrystals display sharp peaks at arbitrarily small momenta.
Important manifestations of this non-trivial order on all length scales include the absence of universal power-law scaling near criticality~\cite{szabo_non-power-law_2018} and its application to quantum complexity~\cite{cubitt_undecidability_2015}.
Moreover, quasicrystals exhibit fascinating
phenomena such as phasonic degrees of freedom~\cite{edagawa_high_2000, freedman_wave_2006, steurer_quasicrystals:_2018}.
To date, quasicrystals have been extensively studied in condensed matter and material science~\cite{shechtman_metallic_1984, edagawa_high_2000, steurer_twenty_2004, barber_aperiodic_2009, steurer_quasicrystals:_2018},
in photonic structures~\cite{freedman_wave_2006, chan_photonic_1998, mikhael_archimedean-like_2008, kraus_topological_2012-1, dareau_revealing_2017},
using laser-cooled atoms in the dissipative regime~\cite{guidoni_quasiperiodic_1997, guidoni_atomic_1999}, and very recently in twisted bilayer graphene~\cite{ahn_dirac_2018}.
Quasicrystalline order can even appear spontaneously in dipolar cold-atom systems~\cite{gopalakrishnan_quantum_2013}.

In this work we realise a quasicrystalline potential for ultracold atoms based on an eightfold rotationally symmetric optical lattice, thereby establishing a new experimental platform for the study of quasicrystals.
Optical lattices, i.e.\ standing waves of light, have become a cornerstone in experimental research on quantum many-body physics~\cite{bloch_quantum_2012}.
They offer an ideal environment for examining quasicrystals since optical potentials are free of defects which greatly complicate measurements on quasicrystalline solids~\cite{steurer_quasicrystals:_2018}. 
In addition, we are able to directly impose `forbidden' rotational symmetries, thereby circumventing the elaborate synthesis of stable single crystals~\cite{feuerbacher_synthesis_2006}.
So far, quasiperiodic optical lattices have been used as a proxy for disorder in ultracold quantum gases~\cite{lye_effect_2007, roati_anderson_2008, derrico_observation_2014, schreiber_observation_2015, bordia_probing_2017}, but the intriguing properties of quasicrystalline order have remained unexplored.
Here we use a Bose-Einstein condensate of $^{39}$K atoms to probe a quasicrystalline optical lattice in a matter-wave diffraction experiment, namely Kapitza-Dirac scattering~\cite{gupta_coherent_2001}.
This allows us to observe a self-similar diffraction pattern, similar to those obtained by Shechtman et al.\ using electron diffraction~\cite{shechtman_metallic_1984} in their original discovery of quasicrystals.
Additionally, we investigate the diffraction dynamics which at short times constitutes a continuous-time quantum walk on a four-dimensional (4D) homogeneous tight-binding lattice.
Confined synthetic dimensions, which can be created by employing the discrete hyperfine states of atoms, already play an important role in quantum simulation~\cite{celi_synthetic_2014, mancini_observation_2015, price_four-dimensional_2015}.
Our measurements demonstrate the potential of quasicrystalline optical lattices to be used for the simulation of extended higher dimensions.

\begin{figure*}[t!]
\begin{center}
\includegraphics[width = 1.0\textwidth]{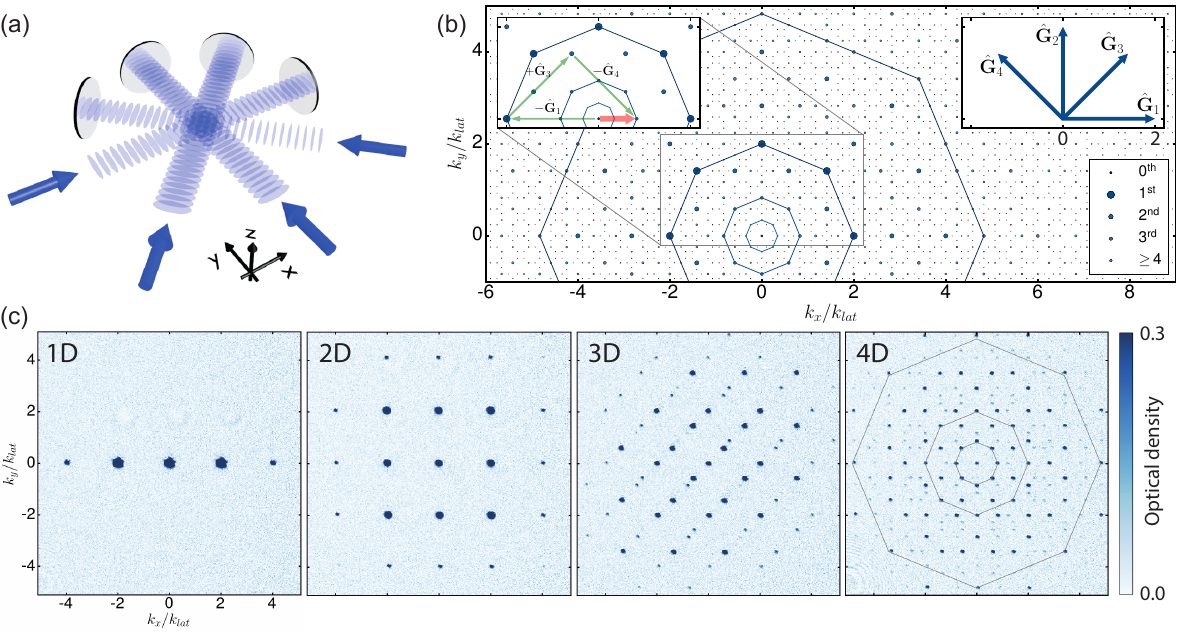}
\caption{(a) Schematic of the eightfold optical lattice formed by superimposing four independent 1D lattices.
  (b) Fractal momentum space structure.
  The first 15 orders of possible diffraction peaks are shown.
  They are constructed by iteratively adding or subtracting one of the four reciprocal lattice vectors $\ki$ (inset on the right) to the peaks in the previous order, starting with $\vec{k} = (0,0)$.
  This results in a fractal structure, whose self-similarity is illustrated by a sequence of octagons, which are each scaled by the silver mean $\sm$ relative to the next.
  The left inset shows one inflation step (see text).
  (c) Raw time-of-flight images resulting from 
  four different lattice configurations at fixed lattice pulse duration ($t = 3.5\,\mu\textrm{s}$).
  Using just one of the lattice axes results in a regular 1D simple-cubic lattice characterized by $\kon$; adding the perpendicular lattice creates a regular 2D square lattice with $\kon$ and $\ktw$.
  By adding the first diagonal lattice we obtain a regular array of quasiperiodic 1D lattices.
  These are characterised by a dense sets of momentum states along $\kth$ whereas the direction perpendicular to $\kth$ remains periodic (labelled 3D).
  Finally, using all four axes we create the 2D quasicrystal (labelled 4D) whose self-similarity is illustrated by the octagons.
}
\label{fig:1}
\end{center}
\end{figure*}

We create the 2D quasicrystalline potential using a planar arrangement of four mutually incoherent 1D optical lattices, each formed by retro-reflecting a single-frequency laser beam, as shown schematically in \mbox{Fig.~\ref{fig:1} (a)}.
The angle between two neighbouring lattice axes is $45(1)^{\circ}$, similar to the setup proposed in ref.~\cite{jagannathan_eightfold_2013} (see also Refs.~\cite{sanchez-palencia_bose-einstein_2005, cetoli_loss_2013}), thereby imposing a global eightfold rotational symmetry in close analogy to the octagonal tiling.
The right inset of Fig.~\ref{fig:1} (b) shows the reciprocal lattice vectors $\kon$, $\ktw$, $\kth$, and $\kfo$ of the four 1D lattices.
In contrast to a periodic lattice the combination of several $\ki$ here may give rise to new, smaller momentum scales, as shown the left inset of Fig.~\ref{fig:1} (b);
for example, the combination $- \kon + \kth - \kfo$ results in a new $k$-vector (red arrow) that is shorter than the original $\kon$ by a factor of $\sm$ (the silver mean).
This process can be repeated ad infinitum and results in a self-similar fractal structure containing arbitrarily small $k$-vectors, giving rise to the sequence of octagons in Fig.~\ref{fig:1} (b).
Consequently, it is impossible to assign a maximum characteristic length to this quasicrystal, heralding the presence of structure on all scales.
The set of momenta that are reachable from $\vec{k}_0=(0,0)$ by combining the $\ki$ is dense in the $k_x,k_y$-plane and any element $\vec{G}$ of this set is determined by four integers $(i,j,l,n) \in \mathbb{Z}^4$ as
\begin{equation}\label{eqn:integer-representation}
  \vec{G}=i\kon+j\ktw+l\kth+n\kfo \; \text{.}
\end{equation}
While physical momentum remains two-dimensional, all four integers are nonetheless required to describe a given $\vec{G}$, since $\cos(45^{\circ})=\sin(45^{\circ})= 1/\sqrt{2}$ is irrational and hence incommensurable with unity.
In fact, Fig.~\ref{fig:1} (b) can be viewed as an incommensurate projection of a 4D simple-cubic {`parent'} lattice to the 2D plane, similar to the `cut-and-project' scheme for constructing the octagonal tiling, {starting from} $\mathbb{Z}^4$~\cite{baake_aperiodic_2013}.
By using fewer than four lattice beams we can control the dimensionality of the parent lattice and reduce  $\mathbb{Z}^4$ to $\mathbb{Z}^D$ with $D \in \{1,2,3,4\}$.

\begin{figure*}[t!]
\begin{center}
\includegraphics[width = 1.0\textwidth]{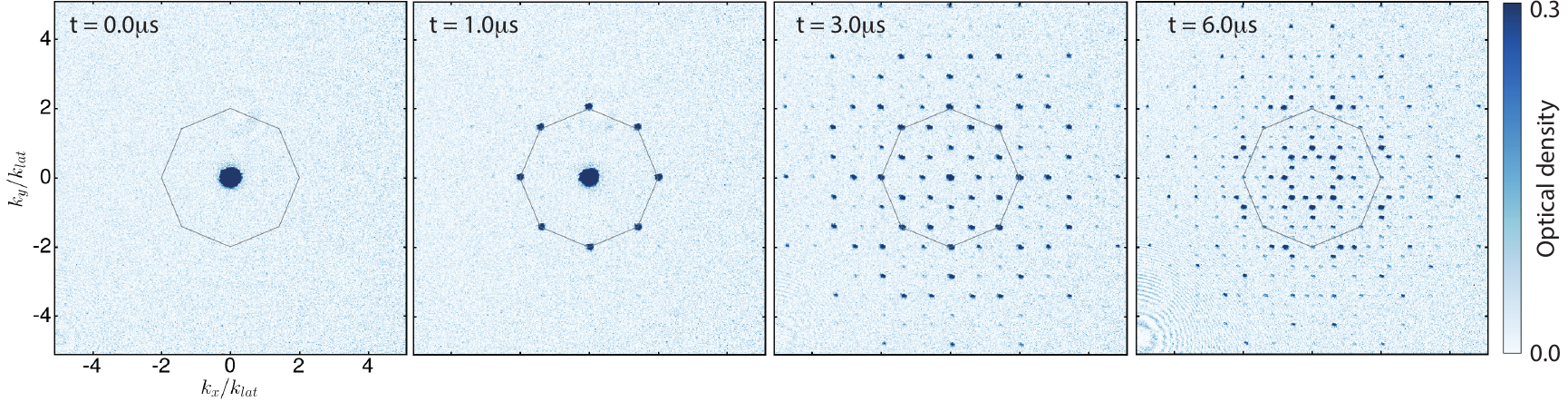}

\caption{Dynamics of Kapitza-Dirac diffraction in the quasicrystalline optical lattice.
The figure shows raw absorption images for four different lattice pulse durations. After $1\mu$s, only the first diffraction order has been populated, while longer pulses lead to populations in successively higher orders as the atoms perform a quantum walk on the fractal momentum structure.
Black octagons with a circumradius of $|\ki|=2\klat$ illustrate the fundamental momentum scale due to two-photon processes.
}
\label{fig:2}
\end{center}
\end{figure*}

The experimental sequence starts with the preparation of an almost pure Bose-Einstein condensate of $^{39}$K atoms in a crossed-beam dipole trap~\footnote{See Supplemental Material, which includes Refs.~\cite{dieckmann_two-dimensional_1998, ketterle_high_1993, greiner_magnetic_2001, simoni_near-threshold_2008, chikkatur_suppression_2000, greiner_exploring_2001}, for further experimental details, theoretical calculations, and data analysis.}.
\nocite{dieckmann_two-dimensional_1998, ketterle_high_1993, greiner_magnetic_2001, simoni_near-threshold_2008, chikkatur_suppression_2000, greiner_exploring_2001}
Using the Feshbach resonance centred at $402.70(3)\,\textrm{G}$~\cite{fletcher_two-_2017} we tune the contact interaction to zero just before we release the condensate from the trap.
Then we immediately expose it to the optical lattice for a rectangular pulse of duration $t$.
During this pulse, atoms in the condensate can undergo several stimulated two-photon scattering events (Kapitza-Dirac scattering~\cite{gupta_coherent_2001}), which scatter photons from one lattice beam into its counterpropagating partner and transfer quantized momenta of $\pm2\hbar\klat$, where $\hbar\klat$ is the momentum of a lattice photon and $|\ki|=2\klat$.
The lattice wavelength $\llat = 2\pi/\klat = 726\,\textrm{nm}$ is far detuned from the $D$-lines in $^{39}$K, ensuring that single-photon processes are completely suppressed.
Throughout this work, the lattice depth of each individual axis is $14.6(2)\Erec$, with $\Erec = h^2/(2 m \llat^2)$ denoting the recoil energy, $m$ being the atomic mass and $h$ being Planck's constant.
Finally, we record the momentum distribution of the atomic cloud by taking an absorption image after $33\,\textrm{ms}$ time-of-flight~\cite{Note1}.

In a first experiment we fix the lattice pulse duration at $t = 3.5\, \mu\textrm{s}$ and vary the number of lattice beams, as shown in Fig.~\ref{fig:1}~(c).
Starting from the single-axis (1D) case, we subsequently add lattice axes,
finally completing the eightfold symmetric case (4D), representing the quasicrystalline structure with its striking self-similarity under ($\sm$) scaling.

\begin{figure}[htbp]
\begin{center}
  \includegraphics[width = 0.5\textwidth]{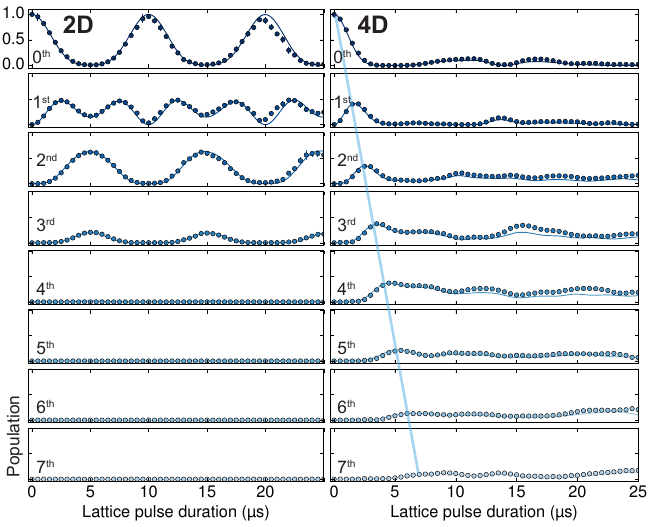}
\caption{Kapitza-Dirac diffraction dynamics in a periodic (2D) and quasicrystalline (4D) lattice.
  The normalized populations (coloured dots) of the condensate (0\textsuperscript{th} order) and the first seven diffraction orders are plotted against pulse duration,
  together with the numerical solution to the Schr{\"o}dinger equation (lines).
  The periodic case (2D) is oscillatory as kinetic energy limits the accessible momenta.
  In contrast, the quasicrystalline lattice (4D) contains a fractal set of $k$-states, c.f.\ Fig.\ref{fig:1} (b), enabling the population of higher and higher orders {without kinetic energy penalty}.
  Correspondingly, the expansion carries on linearly, indicated by the light blue `wave front' as a guide to the eye.
  Error bars denote the standard deviations from five realisations of the experiment, and are typically smaller than symbol size.
}
\label{fig:data-sim}
\end{center}
\end{figure}

The diffraction dynamics offers additional signatures of the fractal nature of the eightfold optical lattice:
during the lattice pulse the condensate explores reciprocal space in discrete steps of $\pm \ki$, leading to profoundly distinct behaviours in the periodic (2D) and in the quasicrystalline case (4D).
Fig.~\ref{fig:2} shows absorption images for four different values of pulse duration $t$ in the latter configuration, illustrating the occupation of more and more closely spaced momenta.
Using individual fits~\cite{Note1} we extract the number of atoms in every $k$-state up to the seventh diffraction order, i.e.\ those momenta reachable by seven or fewer two-photon scattering events.
In all cases, high momentum states are inaccessible, as the corresponding two-photon transitions become off-resonant due to kinetic energy.
Therefore, in the 2D simple cubic lattice (Fig.~\ref{fig:data-sim} on the left) the total number of accessible states is limited and the dynamics is oscillatory, reminiscent of a simple harmonic oscillator.
In the quasicrystalline case (4D, right of Fig.~\ref{fig:data-sim}), in contrast, the diffraction dynamics is non-oscillatory:
due to the fractal momentum space structure, the atoms can access states in ever higher diffraction orders that correspond to ever smaller momenta. 
As a consequence, large parts of the population propagate ballistically to progressively higher orders, as illustrated by the light blue `light cone'.
Our data agrees excellently with exact numerical solutions (lines in Fig.~\ref{fig:data-sim}) of the single-particle time-dependent Schr{\"o}dinger equation in momentum basis~\cite{Note1}.

\begin{figure}[htbp]
\begin{center}
  \includegraphics[width = 0.4\textwidth]{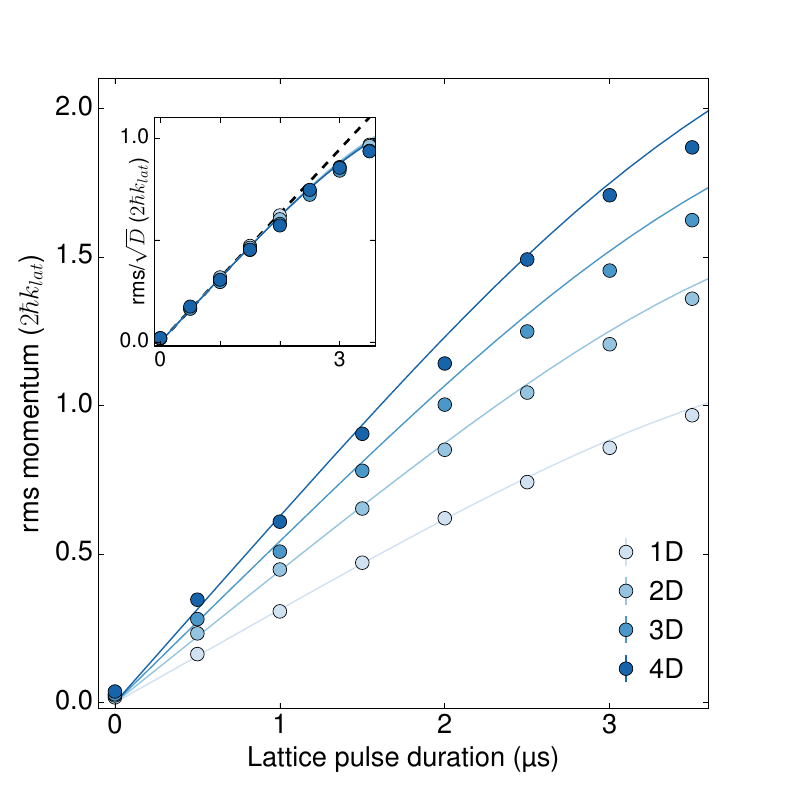}
  \caption{
  Continuous-time quantum walk in $D$ dimensions, where $D$ is controlled by the number of lattice beams.
  Dots represent the measured root-mean-square momentum (see text), while lines represent numerical solutions to the full Schr{\"o}dinger equation.
  The inset shows the same data, but scaled by $\sqrt{D}$.
  Here the dashed line represents the expansion dynamics of a continuous-time quantum walk on a homogeneous $D$-dimensional tight-binding lattice.
  The $\sqrt{D}$ scaling (Eq.~S13 
  in the Supplemental Material) is a direct consequence of the separability of hypercubic lattices.
  Deviations from the linear behaviour at later times are due to kinetic energy, and the lines would differ from each other at long times~\cite{Note1}.
  Error bars denote standard deviations from five identical realisations of the experiment.
}
\label{fig:quantumwalk}
\end{center}
\end{figure}

In the regime of short pulses, the Fourier limit ensures that kinetic energy can be neglected for all dimensions and the discrete momentum space structure can be seen as a homogeneous tight-binding lattice~\cite{gadway_atom-optics_2015, dadras_quantum_2018}.
A hopping event in this effective lattice corresponds to a two-photon scattering event and connects momenta differing by $\pm\hbar\ki$. 
In this picture, the diffraction dynamics is equivalent to the expansion of initially localized particles in this synthetic lattice and gives rise to a continuous-time quantum walk with its characteristic light-cone-like propagation~\cite{weitenberg_single-spin_2011, schneider_fermionic_2012, preiss_strongly_2015}.
For a hypercubic lattice in $D$ dimensions, the separability of the tight-binding dispersion relation leads to an average group velocity proportional to $\sqrt{D}$~\cite{Note1}.
Due to the correspondence between the number of lattice beams and the dimension of the resulting tight-binding hamiltonian, we are able to extend the dynamics to up to four dimensions.
Using the appropriate form of Eq.~\ref{eqn:integer-representation} in $\mathbb{Z}^D$, we extract the effective root-mean-square momentum in $D$ dimensions, e.g.\ $\sqrt{\langle i^2+j^2\rangle}$ in the 2D case and $\sqrt{\langle i^2+j^2+l^2+n^2\rangle}$ in the 4D case, from the individual populations of all diffraction peaks,  and find excellent agreement between the measurements and the analytic result $v_p\propto\sqrt{D}$~\cite{Note1}, as shown in Fig.~\ref{fig:quantumwalk}.
The departure from linear behaviour at longer times is due to kinetic energy and is captured well by the exact numerical solution to the Schr{\"o}dinger equation (solid lines in Fig.~\ref{fig:quantumwalk}). The extent of the linear region is controlled by the lattice depth.
For even longer times, kinetic energy enforces fundamentally different behaviours for periodic and quasicrystalline lattices, as shown in Fig.~\ref{fig:data-sim} (and in Fig.~S3 
in the Supplemental Material).

In conclusion, we have realised a quasicrystalline potential for ultracold atoms, which can facilitate the creation of ever more complex many-body systems~\cite{cubitt_undecidability_2015} and novel phases~\cite{rey_quantum_2006}.
By observing the occupation of successively closer-spaced momenta, we were able to confirm its self-similar fractal structure in momentum space.
In addition, we experimentally verified the fundamentally different diffraction dynamics between periodic and quasicrystalline potentials, in excellent agreement with theory.
Finally, we demonstrated the ability to simulate tight-binding models in one to four dimensions, by observing the light-cone-like spreading of particles in reciprocal space.
On the one hand, these measurements pave the way for more elaborate quantum simulations in four dimensions, including topological effects and charge pumps~\cite{lohse_exploring_2018, kraus_four-dimensional_2013}. 
On the other hand, quasicrystalline potentials enable experimental studies of novel quantum phenomena that have been predicted for quasicrystals, such as non-power-law criticality~\cite{szabo_non-power-law_2018}, topological edge states~\cite{lang_edge_2012, matsuda_topological_2014, singh_fibonacci_2015}, and spiral holonomies~\cite{spurrier_semiclassical_2018}.
Finally, our system will provide unprecedented access to transport and localisation properties of quasicrystals, thereby addressing fundamental questions about the relation between quasiperiodic order and randomness~\cite{khemani_two_2017} and extending studies of many-body localisation and Bose glasses to two dimensions~\cite{derrico_observation_2014, schreiber_observation_2015, meldgin_probing_2016, soyler_phase_2011}.

\appendix

%

\section{Acknowledgments}

\begin{acknowledgments}
  We would like to thank Oliver Brix, Michael H{\"o}se, Max Melchner, and Hendrik von Raven for assistance during the construction of the experiment. We are grateful to Dmytro Bondarenko and Anuradha Jagannathan, as well as Zoran Hadzibabic, Rob Smith, and their team for helpful discussions. This work was partly funded by the European Commision ERC starting grant \mbox{QUASICRYSTAL} and the EPSRC Programme Grant \mbox{DesOEQ} (EP/P009565/1).

\end{acknowledgments}


\newpage
\cleardoublepage

\setcounter{figure}{0} 
\setcounter{equation}{0} 

\renewcommand\theequation{S\arabic{equation}} 
\renewcommand\thefigure{S\arabic{figure}} 

\section{Supplemental Materials}

\subsection{Experimental setup}
The Bose-Einstein condensate of $^{39}$K is produced by a combination of laser cooling, sympathetic cooling with $^{87}$Rb, and evaporative cooling as described in brief in the following.
\paragraph{Magneto-optical trap (MOT).} Our initial laser cooling stage consists of simultaneous cooling and trapping of $^{39}$K and $^{87}$Rb in one MOT chamber.
MOT loading is enhanced by using two separate 2D$^+$MOTs~\cite{dieckmann_two-dimensional_1998}, one for each species, leading to initial loading rates of roughly $4\times 10^9$ $^{87}$Rb atoms/s and $2\times 10^9$ $^{39}$K atoms/s.
Note that these numbers have large systematic uncertainties.
After loading $^{87}$Rb for $2.5\,$s and $^{39}$K for $1\,$s, we let most of the $^{39}$K atoms fall into a dark hyperfine level ($^2$S$_{1/2}$, $F = 1$) by reducing the repump laser power, in order to reduce light-assisted collisions in a `temporal' version of the dark-spot MOT~\cite{ketterle_high_1993}.
After a brief molasses cooling stage and optical pumping of $^{87}$Rb into the trapped hyperfine ground state $\left(^2\textrm{S}_{1/2},\,\ket{F = 1, m_F = -1}\right)$, we load on the order of $10^9$ $^{87}$Rb and $7\times 10^8$ $^{39}$K atoms into a magnetic quadrupole trap.
The temperature at this stage is about $80\,\mu \textrm{K}$ for $^{87}$Rb and $150\,\mu \textrm{K}$ for $^{39}$K at a gradient of $100\,\textrm{G}/\textrm{cm}$.

\paragraph{Magnetic transport and forced microwave evaporation.} The combined clouds are transported by successively ramping nineteen pairs of quadrupole coils~\cite{greiner_magnetic_2001} to reach the glass chamber in which all experiments are performed.
In the main quadrupole trap, where we use gradients of up to $300 \textrm{G}/\textrm{cm}$, we perform forced evaporation of $^{87}$Rb using microwave radiation generated by a mixing the output of a commercial fixed-frequency microwave source with a home-built direct-digital-synthesis module.

\paragraph{Dipole trap.} The final cooling stages happen in a crossed-beam dipole trap, at the start of which we capture roughly $9\times 10^6$ $^{87}$Rb atoms at $6.6\,\mu \textrm{K}$ and $5\times 10^6$ $^{39}$K atoms at $10.5\,\mu \textrm{K}$.
Before evaporating in this dipole trap, we perform a simultaneous radio-frequency state transfer for both species from the respective $\ket{F = 1, m_F = -1}$ to the $\ket{F = 1, m_F = 1}$ hyperfine ground state at a small homogeneous magnetic field.
We then apply a homogeneous offset field of approx.\ $317.7(1)\,\textrm{G}$ that is produced by the same magnetic coils that were previously used for the quadrupole trap, after having swapped the polarity of one of the coils using a high-current H bridge.
The error bars for our B-field values are systematic uncertainties in our calibration.
This field corresponds to a positive s-wave scattering length on the order of $10^3\,a_0$ for the inter-species contact interaction between $^{87}$Rb and $^{39}$K~\cite{simoni_near-threshold_2008}, where $a_0$ denotes the Bohr radius.
Subsequently the power in each dipole trap beam is reduced from 7W to 210mW in a 4-second-exponential ramp, at some point during which all $^{87}$Rb atoms have been lost due to gravity.
At this point we change the field to $397.8(1)\,\textrm{G}$, corresponding to roughly $280\,a_0$ for the intra-species Feshbach resonance of $^{39}$K~\cite{fletcher_two-_2017}.
The final trapping frequencies for $^{39}$K in the dipole trap are $2\pi\times15(1)\,\textrm{Hz}$ in the horizontal directions and $2\pi\times84(1)\, \textrm{Hz}$ in the vertical, axial direction.
This sequence yields $1.5(2)\times 10^5$ $^{39}$K atoms in a Bose-Einstein condensate with no discernible thermal fraction.

\paragraph{Optical setup of lattice beams and dipole trap.}
All dipole and lattice beams propagate in the horizontal plane.
The two dipole beams are derived from a single-frequency solid state laser at $1064\,\textrm{nm}$ and are overlapped with each other at an angle of just under $90^{\circ}$ at the position of the atoms.
All lattice and dipole trap beams have elliptical profiles; the dipole beams have a vertical single waist of $60(5)\,\mu\textrm{m}$ and a horizontal single waist of $290(10)\,\mu\textrm{m}$, whereas the lattice beams have single waists of $70(3)\,\mu\textrm{m}$ (vertical) and $160(5)\,\mu\textrm{m}$ (horizontal). 

The lattice beams are derived from one single-frequency Ti:Sa laser at $\llat = 726\,\textrm{nm}$ which is far-detuned with respect to the $D$-lines in $^{39}$K, ensuring suppression of single-photon scattering.
Cross-interferences between lattice axes are avoided by offsetting their frequencies by more than $10\,\textrm{MHz}$ from each other.
Therefore, the corresponding beat notes between the axes oscillate much faster than the atomic kinetic energy scale, given by $\Erec/h = 9.7\,\textrm{kHz}$ for $^{39}$K.

\paragraph{Interaction effects.}
For diffraction experiments that are carried out with finite contact interactions (e.g.\ using $^{87}$Rb) one finds that the time-of-flight images feature pronounced `scattering shells'~\cite{chikkatur_suppression_2000, greiner_exploring_2001} connecting the discrete momentum peaks. These shells appear as characteristic rings on the absorption images and arise from two-body s-wave collisions between parts of the atomic cloud which are moving with respect to each other.
In order to eliminate this effect we tune the contact interaction in $^{39}$K to zero by ramping the magnetic field to a value of $351.5(1)\,\textrm{G}$ just before the optical lattice pulse is applied.
However, atomic clouds of $^{39}$K at vanishing interactions are optically dense enough to absorb essentially all imaging light, preventing any faithful atom number measurement.
Therefore we turn interactions back on (back to the previous value of roughly $280\,a_0$) once the diffraction orders have separated from each other, such that the individual peaks expand and reveal their atom populations.

The stronger mean-field expansion of highly populated peaks results in their widths being larger after time-of-flight. Furthermore, in the chosen colour scale all optical densities above 0.3 are represented in blue.
Therefore, a larger occupation (higher central density) means that the blue area in the plot is larger even for the same physical width of the peak.
These effects are clearly visible in the $ \vec{k} = (0, 0)$ peak in the first panels of Fig.~2.

\subsection{Theoretical model to describe the dynamics of atoms exposed to a short flash of the optical lattice}
The real-space potential of our optical lattice can be written as
\begin{equation}\label{eqn:potential-real}
  \Vd = V_0 \sum_{i = 1}^D \cos^2 \left(\frac{\ki}{2} \cdot \vec{r}\right)  \text{,}
\end{equation}
where $D = 1,2,3,$ or 4 is the number of mutually incoherent lattice beams and $V_0 = 14.6(2)\,\Erec$ is the individual lattice depth.
The reciprocal lattice vectors $\ki$ are defined as
\begin{eqnarray}\label{eqn:latvecs}
  \kon  =  \two{1}{0}\text{,} \quad \ktw  =  \two{0}{1}\text{,}\\[1ex]
  \kth  = \sqrttwo \two{1}{1}\text{,} \quad \kfo  =  \sqrttwo \two{-1}{1} 
\end{eqnarray}
where we have switched to dimensionless units in which $2\klat = 1$ as shown in Fig.~1(b) in the main text (right inset).

\paragraph{Basis in momentum space.}
We can use Eq.~1 in the main text to express any accessible $k$-state as an integer-valued vector $\bi$ in $\mathbb{Z}^D$, where $D$ is the number of active lattice beams.
The $n$th order of this basis corresponds to the $n$th order of diffraction peaks in Fig.~3 in the main text and is defined as the set of all elements $\{\bi\}$ with 
\begin{equation}\label{eqn:generation}
 \sum_{j=1}^D \text{abs}\left(\left[\bi\right]_j\right) = n
\end{equation}
where $\left[\vec{a}\right]_j$ denotes the $j$th component of a vector $\vec{a}$ and $\text{abs}()$ denotes the absolute value.
We will later truncate $\mathbb{Z}^D$ at the $n$th order, meaning we only take into account states that can be reached by at most $n$ two-photon scattering events.

\paragraph{Projection.}

For $D=3$ and $D=4$ the projection matrices for the states $\bi$ onto the $k_x,k_y$--plane are given by
\begin{equation}
    M_3 = \left( {\begin{array}{ccc}
   1 & 0 & \sqrttwo\\
   0 & 1 & \sqrttwo\\
   0 & 0 & 0\\
  \end{array} } \right)
\end{equation}
and
\begin{equation}
    M_4 = \left( {\begin{array}{cccc}
   1 & 0 & \sqrttwo & -\sqrttwo\\
   0 & 1 & \sqrttwo & \sqrttwo\\
   0 & 0 & 0 & 0\\
   0 & 0 & 0 & 0\\
  \end{array} } \right) \text{,}
\end{equation}
respectively. The third and fourth dimensions are projected to the in-plane diagonals with respect to the first two dimensions.
The first two rows of $M_3$ and $M_4$ are simply given by the $\ki$ defined in Eq.~\ref{eqn:latvecs}.
Fig.~1(b) in the main text shows elements of the basis of order 15 for $D=4$, projected onto the $k_x, k_y$--plane using the matrix $M_4$.
For $D=1(2)$ the matrices $M_1(M_2)$ are trivial because here the dimension of $\bi$ is the same as the physical dimension.

\paragraph{Hamiltonian in momentum space.}
Having constructed the basis we can write down the hamiltonian $\mathcal{H}^D$ in momentum space
\begin{eqnarray}\label{eqn:hamiltonian}
  \mathcal{H}^D_{i,j} = \begin{cases}
    V_0/4  \qquad &\text{for}\quad \vert\bi - \bj\vert = 1\\
    4\Erec \times \vert M_D \cdot \bi\vert^2 + V_0/2 \qquad &\text{for}\quad i = j\\
    0 &\text{otherwise}
  \end{cases}
\end{eqnarray}
Here the norm of a $D$-dimensional vector $\vec{a}$ is given by
\begin{equation}\label{eqn:norm}
\vert\vec{a}\vert = \sqrt{\sum_{j=1}^D \left(\left[a\right]_j\right)^2}\quad\text{.}
\end{equation}
The non-zero off-diagonal elements correspond to transition elements for stimulated two-photon scattering events, where atoms scatter photons from one lattice beam into its counterpropagating partner. These transitions connect discrete momentum states separated by $\pm \ki$ and effectively realise a tight-binding hamiltonian in momentum space~\cite{gadway_atom-optics_2015}.
The matrix elements on the diagonal are given by the kinetic energy term, where the prefactor $2^2\Erec$ arises from the momentum scale $2\hbar\klat$ of the individual lattices and a constant offset of $V_0/2$ which arises from the $k = 0$ Fourier component of $\Vd$.
\paragraph{Truncation of basis.}
In principle, this hamiltonian is infinite-dimensional and, consequently, we need to make it numerically tractable by truncating it.
Since the experiment starts with a pure condensate in the $\ket{\hbar \vec{k} = 0}$ state, and we apply only short pulses of lattice light, it is sensible to work with a basis of order $n=11$.
This can be justified a posteriori since even for the longest applied lattice pulses our simulation (using a basis of order 15) shows that the orders $n>11$ get populated by less than 15 per cent.

\paragraph{Time-evolution.}
In order to simulate the time-evolution of our diffraction experiment (solid lines in Figs.~3 and~4 in the main text) we numerically integrate the time-dependent Schr{\"o}dinger equation including the hamiltonian matrix $\mathcal{H}^D_{i,j}$ in the truncated momentum basis using the \verb scipy.integrate \ library for integrating differential equations.

\paragraph{Lattice depth calibration.}
In the one- and two-dimensional cases, higher diffraction orders do not get populated and the solution to the Schr{\"o}dinger equation with a truncated basis becomes exact. By carrying out 1D diffraction experiments with each of the four individual lattice axes and comparing the summed populations with the theoretical prediction (as in Fig.~3 in the main text) we can calibrate our individual lattice depths. These calibrations are consistent with an independent calibration using amplitude modulation and parametric heating.

However, experiments involving several lattice beams and long pulse durations start to be sensitive to residual relative mismatches in the lattice depths (typically less than $5 \%$), resulting in different `oscillation frequencies' in different directions in the periodic lattices.
This effect can be seen in Fig.~3 (2D) in the main text: the second revival of the zeroth order momentum peak does not reach the theoretically predicted maximum since the oscillation frequencies in the two orthogonal lattice directions are slightly different.
In principle, the lattice depth could be calibrated to an even finer degree to avoid this effect.
However, we find that for the 4D case and long lattice pulses the absorption image analysis is already limited by the finite signal-to-noise ratio of our detection method (see below).
Therefore, a finer lattice depth calibration would not yield cleaner results and we limit our observation to lattice pulse durations $t\leq25\mu\textrm{s}$.

\subsection{Absorption image analysis}

\begin{figure}[htbp]
\begin{center}
  \includegraphics[width = 0.5\textwidth]{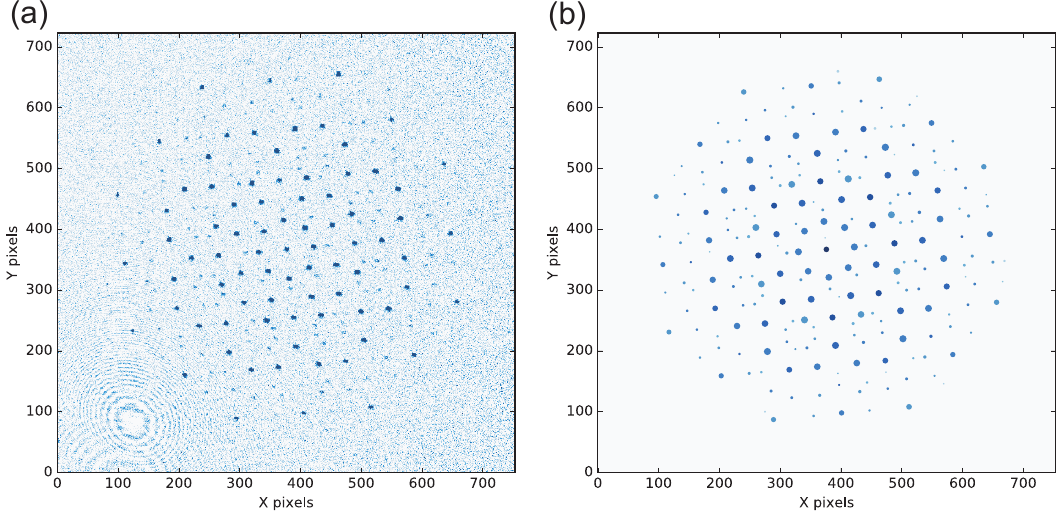}
  \caption{Simplified example of the population count. (a) The raw absorption image ($D = 4, t = 3\mu\textrm{s}$). At each calculated peak position with  $n\leq 7$ we perform an individual fit to a 2D Thomas-Fermi profile. The resulting populations $p_i(t)$ are depicted in (b) by the area of the circles.
}
\label{fig:bins}
\end{center}
\end{figure}

\paragraph{Populations in diffraction peaks.}
First, we determine the position of the condensate, find the angle of one lattice axis relative to the camera axes, and calibrate the magnification using reference images showing only zeroth and first order diffraction peaks.
With this information we can calculate the expected position of each momentum peak.
Around each calculated peak position we perform an individual fit to a 2D Thomas-Fermi profile (a paraboloid) in a square bin of $28\times28$ pixels ($56\times56$ for the central condensate).
In order to mitigate effects of imaging saturation, the fit ignores pixels with optical densities above $2.0$.
The corresponding atom population of each basis element $p_i(t)$ is proportional to the integrated Thomas-Fermi profile, as shown as an example in Fig.~\ref{fig:bins}.
If this population value is below $0.04\%$ of the total population we ignore it in order to avoid counting spurious populations in high diffraction orders, which would otherwise dominate the rms.
For Fig.~3 in the main text we sum all populations in one diffraction order.

\paragraph{Root-mean-square extraction.}
We calculate the root-mean-square momentum in $D$ dimensions as a function of time as
\begin{equation}\label{eqn:rms4d}
\sqrt{\sum_{i} \frac{p_i(t)}{\sum_j p_j(t)} \vert\bi\vert^2} \qquad \text{,}
\end{equation}
where $p_i(t)$ are the populations in each diffraction peak $\bi$ at a given time $t$. The sums go over all elements of a basis.

\begin{figure}[htbp]
\begin{center}
  \includegraphics[width = 0.5\textwidth]{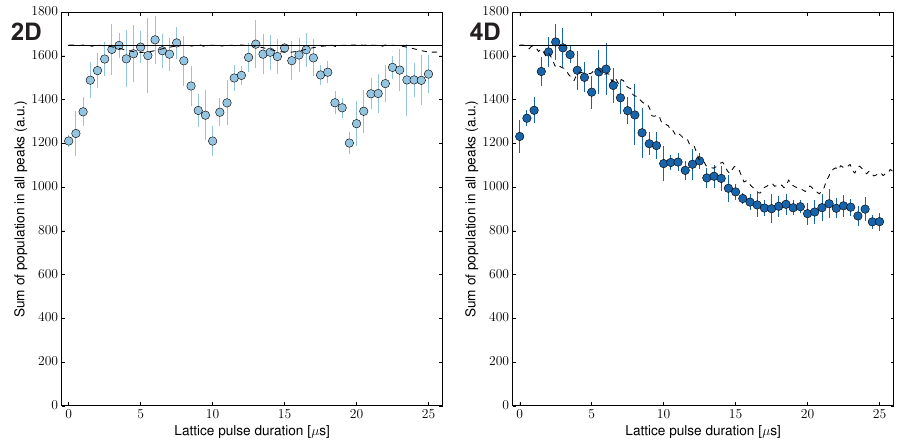}
  \caption{Total detected population (a.u.) in all diffraction peaks for the 2D and 4D situations.
  The solid black lines correspond to the total population and the dashed lines are the simulated populations in peaks up to seventh order and additionally account for peaks that fall below the cutoff.
  When single peaks (such as the central condensate) are strongly populated we systematically detect too few atoms due to imaging saturation and the finite signal-to-noise ratio of the camera.
  This effect is reflected by the apparent rise in total detected population during the first few $\mu$s. It also explains the subsequent `dips' in atom number in the 2D case.
}
\label{fig:numbers}
\end{center}
\end{figure}

\paragraph{Total population.}
Figure~\ref{fig:numbers} shows the total detected population $\sum_j p_j(t)$ summed over all diffraction peaks for the cases $D=2,\,4$.
For $D>2$, the detected population is reduced for longer lattice pulses since more and more peaks are weakly populated and fall below the cutoff.
In addition, we underestimate the population of very highly populated peaks, such as the initial condensate, due to the finite signal-to-noise of the camera.

\subsection{Group velocity estimate}

In this section we will derive the $\propto \sqrt{D}$ scaling of the rms expansion in a homogeneous tight-binding lattice of dimension $D$. This description is valid in the limit of short pulse durations where the kinetic energy terms in Eq.~\ref{eqn:hamiltonian} can be ignored.
Let us first consider a homogeneous tight-binding model on a $D$-dimensional hypercubic lattice with spacing $a$ and hopping matrix element $J$.
The dispersion relation for an eigenstate with quasimomentum $\vec{q}$ is given by
\begin{equation}
  E(\vec{q}) = -2J \sum_{i = 1}^D \cos(a q_i) \quad \text{.}
\end{equation}
Correspondingly, the individual components of its group velocity are given by
\begin{equation}
  v_i(\vec{q}) = \frac{1}{\hbar} \frac{\partial E(\vec{q})}{\partial q_i} = \frac{2Ja}{\hbar} \sin(aq_i) \quad \text{.}
\end{equation}
For a given Wannier state that is initially localized to one lattice site all Bloch waves are equally populated, leading to an average root-mean-square group velocity
\begin{equation}\label{eqn:v}
  \sqrt{\overline{\vec{v}^2}} = \frac{2Ja}{\hbar} \sqrt{\sum_{i=1}^D \overline{\sin^2(aq_i)}} = \frac{2Ja}{\hbar}\sqrt{\frac{D}{2}} \quad \text{.}
\end{equation}
Now we switch from a real-space model to our tight-binding model in momentum space.
This corresponds to replacing the lattice spacing $a$ and the hopping matrix element $J$ in Eq.~\ref{eqn:v} with $2\hbar\klat$ and $V_0/4$, respectively, resulting in
\begin{equation}
  v_p \equiv \frac{\sqrt{\overline{\vec{v}^2_p}}}{2\hbar\klat} = \frac{V_0}{2\hbar} \sqrt{\frac{D}{2}} \quad \text{.}
  \label{eqn:vk}
\end{equation}
Here $v_p$ is the group velocity in momentum space in units of momentum ($2\hbar\klat$) per unit time.

\begin{figure}[htbp]
\begin{center}
  \includegraphics[width = 0.5\textwidth]{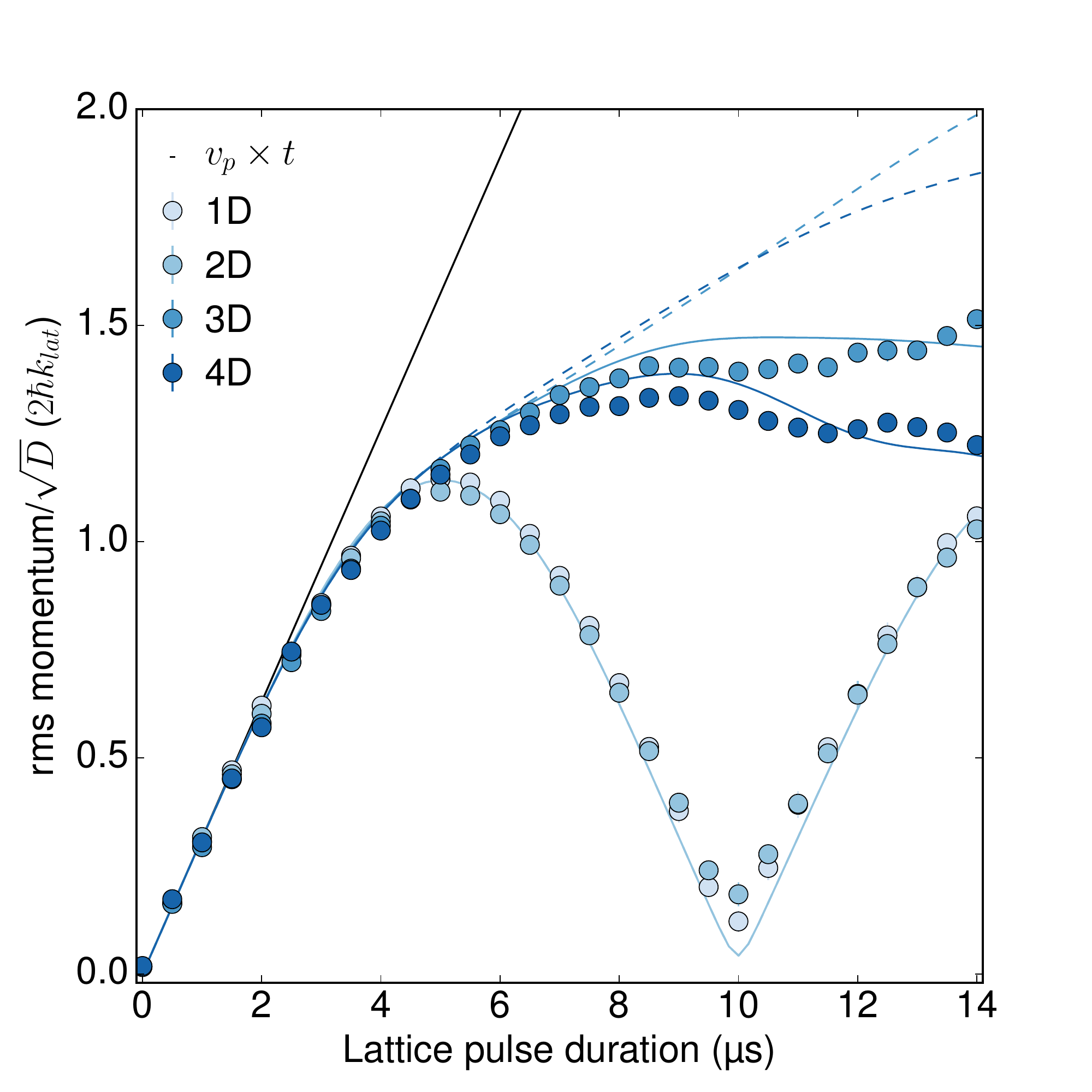}
  \caption{Root-mean-square momentum for longer times, scaled by a factor $\sqrt{D}$ as in Fig.~4 (inset) in the main text.
  The dashed black line is the analytic result (Eq.~\ref{eqn:vk}) in a $D$-dimensional homogeneous tight-binding model.
  The dashed blue lines are the solutions to the time-dependent Schr{\"o}dinger equation, using all states with $n\leq11$.
  If we take into account that we only detect atoms up the seventh diffraction order, the expansion is reduced to slightly lower momentum values (solid lines).
}
\label{fig:rms-long}
\end{center}
\end{figure}

The assumption of neglecting kinetic energy breaks down for longer lattice pulse durations and the rms expansion begins to deviate from the linear behaviour, as shown in Fig.~\ref{fig:rms-long}.
In periodic potentials (1D and 2D cases) only a few $k$-states are accessible, leading to a revival of the condensate after a certain period.
In the quasiperiodic cases (3D and 4D) there are infinitely many $k$-states within reach and the populations can propagate to successively higher orders without kinetic energy penalty.

\clearpage
\end{document}